\documentclass{article}
\usepackage{natbib}
\usepackage[british]{babel}
\usepackage[varg]{txfonts}
\usepackage{rotating}
\usepackage{dcolumn}

\newcommand\apj{\rmfamily{ApJ}}%
\newcommand\apjl{\rmfamily{ApJ}}%
\newcommand\apjs{\rmfamily{ApJS}}%
\newcommand\apss{\rmfamily{Ap\&SS}}%
\newcommand\aap{\rmfamily{A\&A}}%
\newcommand\solphys{\rmfamily{Sol.~Phys.}}%
\newcommand\ssr{\rmfamily{Space~Sci.~Rev.}}%

\begin{document}

\title{Quenching of the alpha effect in the Sun -- what observations are telling us} 
\author{R.~H.~Cameron\thanks{email:\texttt{cameron@mps.mpg.de}} \\
       \small{Max-Planck-Institut f\"ur Sonnensystemforschung, Katlenburg-Lindau, Germany}}


\date{Received \today}

\maketitle
\label{firstpage}

\begin{abstract}

The Babcock-Leighton type of dynamo has received recent support in terms of the 
discovery in the observational records of systematic cycle-to-cycle variations in the tilt 
angle of sunspot groups. It has been proposed that these variations might be the consequence 
of the observed inflow into the active region belt. Furthermore simulations
have shown that such inflows restrict the creation of net poloidal flux, in effect acting
to quench the alpha effect associated with the Coriolis force acting on rising 
flux tubes. In this paper we expand on these ideas.

\end{abstract}


\section{Introduction}\label{s:intro}

The Babcock-Leighton type of dynamo has received recent support:  following the
discovery by \cite{Dasi-Espuig10}  of systematic cycle-to-cycle variations in the tilt
angle of observed sunspot groups, \cite{Cameron10} used the Surface Flux Transport Model
to show that the polar field at the end of cycle $n$ is correlated with the strength of 
cycle n+1 for cycles 13-21 \citep{Cameron10}. This strengthens the result from direct 
observation of the field which only covered cycles 20-23.  
The observations thus strongly suggest the solar dynamo is of the Babcock-Leighton type.

The Babcock-Leighton dynamo is essentially linear in its description, and does not comment
on what limits the strength of the different cycles. We can easily see that it must involve
a modification of the flow field, so the
question becomes one of where, and on what scales, does the magnetic field affect the flow.
In the traditional alpha quenching of mean-field dynamos the flows are modified on small 
scales and in the region where there is toroidal field.  
\cite{Kitchatinov11}, however, note that in Babcock-Leighton dynamos the alpha effect occurs
in a different location (in the bulk of the convection zone or near the solar surface) to where 
the strong toroidal fields are stored (the bottom of the convection zone). They argue that
traditional alpha quenching does not apply in this case.
The rest of this paper outline an observationally suggested mechanism for
quenching the alpha effect associated with the Coriolis force acting on rising flux tubes.

\section{Schematic outline of the Babcock-Leighton dynamo and the non-linear feedback.} 

The basic Babcock-Leighton dynamo mechanism is illustrated by the blue boxes
and brown arrows in Figure~1. From top to bottom, we begin with toroidal
flux near the bottom of the convection zone, then an instability \citep{Parker66,Spruit82},
or series of instabilities (eg. the instability described in  \citeauthor{Rempel01}, \citeyear{Rempel01}, 
followed by that in
\citeauthor{Parker66}, \citeyear{Parker66}) leads to a flux tube buoyantly rising to
the surface. The Coriolis force is thought to tilt the tube during its rise 
\citep{DSilva93}.  As a conseuence of the twist, when the flux emerges at the surface in the 
form of sunspots, the leading spots are slightly closer to the equator than the following spots.
Over a period of about a month the spots breaks up into faculae, which is acted on by
the systematic meridional flow and differential rotation, as well as the random (on the
scale of the magnetic field) granular and supergranular flows. 
The random flows
can be treated as a type of diffusion on the timescales relevant for the solar cycle. 
Whilst most of the flux is advected towards the nearest pole, some diffuses across the equator.
Because the leading polarity flux is slightly closer to the equator than the following polarity,
the equator-crossing flux is preferentially of the leading spot polarity in both 
hemispheres and hence in each cycle there is net
flux reaching the pole. The dynamo model relies on this imbalance being sufficient to reverse
the polar field. The polar field is then associated with poloidal flux threading
the sun which then gets wound up by differential rotation to provide the new toroidal 
flux (of the opposite sign to that previously present) at the base of the convection zone.
This closes the loop (of half a 22-year magnetic cycle).

\begin{figure}
\centerline{\includegraphics[width=10cm]{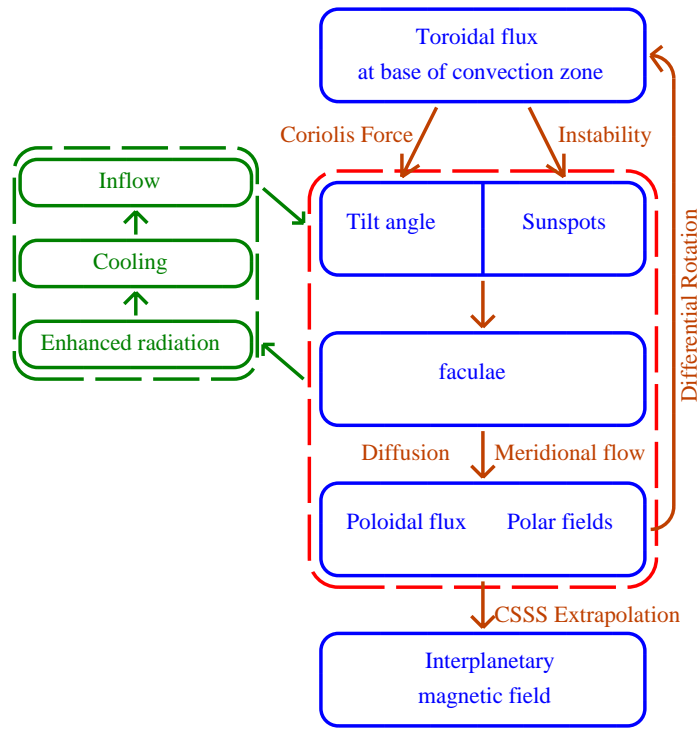}}
\caption{Schematic illustration of how the surface flux transport model (inside the
red dashed box) is embedded in the Babcock-Leighton dynamo model (blue boxes) and
how the enhanced radiation associated with facular regions can lead to what is essentially
alpha quenching.}
\end{figure}

The surface part of the model is enclosed in the red dashed line in Figure 1, and it is this part
which is also dealt with by the Surface Flux Transport Model (SFTM). The historical record of 
sunspot emergence and cycle-averaged tilt angles can be fed in as inputs to the SFTM and the
surface field therefrom extrapolated outwards using the Current Sheet Source Surface model
\citep{Zhao95}. 
The 
resulting polar field and interplanetary field are then retrieved as outputs.  
\cite{Cameron10} used this approach, found that the resulting interplanetary field matches
that inferred from measurements of the geomagnetic-$aa$ index, and that the polar
fields were well correlated with the activity level of the next cycle. Important
in the current context is that the polar field is essentially proportional to the 
activity level of the next cycle, with the linear fit passing close to the origin.

The fact that the relationship between the polar field and the activity of the next 
cycle is linear  allows us to localize the saturation mechanism
in the conceptual illustration. 
Because the polar field of cycle $n$ is linearly related to the activity of cycle $n+1$ it
follows that the winding up of the poloidal field by differential rotation is, in the Sun,
a linear process. Similarly the condition for toroidal flux to erupt from the base of the 
convection zone cannot vary strongly from cycle to cycle. 
The nonlinearity responsible for the saturation must
therefore be in one or more of: the tilt angles, the granular/supergranular 
diffusivity or the meridional flow. Since the granular and supergranular properties
do not vary much with the cycle \citep[see the review by][]{Rieutord10} the nonlinearity must be
in some combination of the tilt angle and meridional flow. 

The required types of changes are observed in both the near-surface velocity field
\citep{Haber02} and in the tilt angles of sunspot groups \citep{Dasi-Espuig10}. In fact, the two types
of observations are not independent. As suggested by \cite{Dasi-Espuig10}, the 
observed changes in the tilt angles could be produced by the observed local changes of 
the meridional flow. The change in the meridional flow has itself been explained by
the observed difference in radiance between bright facular regions and 
undisturbed quiet-Sun regions. We can then reconstruct
a likely logic chain for the reduction of the tilt angle. It is sketched in the
green boxes in Figure 1. The bright faculae is associated with an enhanced 
emission,  which cools the plasma in the facular regions slightly
more efficiently then in the surrounding quiet-Sun. This produces a small temperature
difference which drives the observed inflow into the activity belt. The scenario
here was proposed by \cite{Spruit03} and was simulated and compared to the helioseismic
results by \cite{Gizon08}.
The inflow, in turn, reduces the latitudinal separation of the opposite 
polarities of sunspot groups (observed tilt angle reduction). Idealized calculations by
\cite{Jiang10a} show that the effect of the inflows on the poloidal flux is significant. 

Because the inflows are caused by activity, their strength should increase with activity,
and indeed \cite{Cameron10b} showed how the observed timedependence of the inflow can be modeled 
as the integral of local inflows assumed to be proportional to the local field strength. 
The calculations of \cite{Jiang10a} show that the amount of net flux escaping to the poles 
decreases with increasing strength of the inflows. 
The activity-related inflows therefore act to quench the alpha effect associated with 
rising flux tubes: strong cycles have a large amount of surface field, which drives strong inflows
and reduces the latitudinal separation between the opposite polarities in sunspot groups. 
It is worth emphasising that the inflow is the integral of the flows produced by individual
active regions. 
The weak integrated inflow of a weak cycle is less able to affect the tilt angles of sunspot groups
than the strong integrated inflow of a strong cycle. This is so even though the sizes and strengths of 
individual active regions of strong and weak cycles are drawn from the same distribution \citep{Jiang11}.
In this way it differs from some of the other mechanisms which affect the tilt angle of active regions,
such as those studied by \cite{DSilva93} and \cite{Nandy}, in that it explicitly links the change in 
the tilt angles to the properties of the cycle during which they emerge. 
The inflows thus act as a non-local (in the sense that the inflows are driven by other active regions and plage)
alpha quenching mechanism.

In passing we also comment that there is still some debate \citep{Svanda08} over whether the meridional 
inflow has a component which is delocalized with respect to the magnetic activity. However since newly emerging regions
do not avoid existing active regions 
\citep[rather they appear to preferentially emerge in them,][]{Harvey93}
this does not greatly affect the argument.

\section{Conclusion}
Solar dynamo theorists appear to be fortunate: observations suggest that an important
mechanism for quenching the alpha effect associated with rising tubes is present
at the solar surface, where it can be observed. Furthermore, the flows involved are large scale, 
and have been mapped below the surface. Our good fortune is based on the fact that, on the
Sun, faculae are substantially brighter than the quiet-Sun; energetically, the power associated with the 
radiance variations integrated over a cycle has been estimated to be similar to that of the magnetic 
field \citep{Schuessler96} and is in phase with the magnetic activity. This need not be the case for other stars \citep{Beeck11},
and if the net brightness enhancement due to activity  becomes too weak (or becomes a brightness deficit) than
the quenching mechanism we have identified here will not apply. In such cases the expectation is that the
cylces would have larger amplitudes and some other non-linearity would be responsible for the saturate. 
There are indications that stars where the brightness fluctuations are anticorrelated with the cyclic activity 
have stronger field strengths \cite{Radick98}, however such stars differ in other systematic ways and other causes
of the stronger activity are possible. 
Future work will see if testable predictions for a range of other types of stars 
can be made for future asteroseismology missions such as PLATO.


\appendix

\label{lastpage}
\end{document}